\documentclass[12pt,a4paper]{article}

\usepackage[utf8]{inputenc}  
\usepackage[T1]{fontenc}
\usepackage{lmodern}         
\usepackage{amsmath,amssymb,amsthm}
\usepackage{geometry}
\usepackage{hyperref}
\usepackage{graphicx}
\usepackage{float}

\usepackage{authblk}
\usepackage{booktabs}
\usepackage{mathtools}

\usepackage{tabularx}
\newcolumntype{C}{>{\centering\arraybackslash}X}

\hypersetup{
  colorlinks=true,
  linkcolor=blue,
  citecolor=blue,
  urlcolor=blue
}

\title{The Trust-Building Game: A Model for Sustainable Cooperation}
\author{
Madjid Eshaghi Gordji\thanks{Corresponding author. Email: meshaghi@semnan.ac.ir}
\and
Mohamadali Berahman\thanks{Email: mohamadali\_berahman@semnan.ac.ir}
}

\affil{
Faculty of Mathematics, Statistics and Computer Science\\
Semnan University, Semnan, Iran
}

\begin{document}

\maketitle
\begin{abstract}
Trust serves as a fundamental pillar of human interactions, playing a crucial role in economic, social, and political relationships. While traditional models of trust primarily focus on the decision-making of the first player, this paper introduces an innovative approach that shifts the emphasis to the decision-making processes of the second player. The proposed model employs a repetitive structure through which the first player effectively cultivates the second player's trust through consistent and strategic actions. This framework is applicable across various fields, including public policy, marketing, and international relations. It significantly deepens our understanding of the trust-building process and aids in fostering sustainable cooperation in real-world scenarios.
\end{abstract}

\section{Introduction}

In the theoretical literature of game theory, the classical trust game, first introduced by (Berg, Dickhaut et al., 1995) as the investment game, has been widely adopted as a foundational framework for analyzing trust behavior under conditions of uncertainty. In this game, the first player decides how much of their initial endowment to transfer to the second player, without any assurance of reciprocation. The second player then has the option to return part or all of the amount received, typically increased by a multiplier embedded in the game’s mechanism, or to retain the entire sum. This structure has been extensively employed to quantify individuals' willingness to place trust and to evaluate trustworthiness in the absence of enforceable commitments.

Trust is regarded as one of the fundamental pillars of human interaction and plays a central role in shaping social, economic, and political relationships. Accordingly, the investment game has been established as a standard model for analyzing decision-making processes under conditions where social risk and expectations of reciprocity are at play. The significance of this model becomes particularly evident in contexts where explanations based solely on economic rationality fall short in capturing human behavior, thereby necessitating the consideration of psychological, ethical, and social factors.

For instance, meta-analytical investigations such as the study by (Johnson and Mislin, 2012) have demonstrated that experimental design details and cross-cultural differences can significantly influence levels of trust and trustworthiness. Moreover, in an earlier meta-analysis based on metadata from 84 trust game experiments, (Johnson and Mislin, 2011) found that approximately 40\% of the variation in trust and 30\% of the variation in trustworthiness could be attributed to differences in experimental protocols. In addition, they identified significant associations between socio-cultural variables and trust-related outcomes; for example, greater ethnic diversity was associated with lower levels of trustworthiness, and class-based disparities were linked to shifts in players’ trust patterns.
From a psychological perspective, (Charness and Dufwenberg, 2006) examined the influence of non-binding promises in the trust game and found that making a promise—even in the absence of enforcement mechanisms—can significantly affect player behavior. Their results suggest that individuals often attempt to honor their commitments to avoid feelings of guilt, which in turn promotes cooperation and leads to higher rates of investment return. However, (Zhang, Ji et al., 2022) demonstrated that requested promises do not have a statistically significant effect on trust or promise-keeping behavior, and approximately 41\% of participants failed to fulfill their promises even without receiving any monetary compensation. Additionally, (Samson and Kostyszyn, 2015) investigated the role of cognitive load and found that participants exhibited significantly lower trust and more impulsive behavior when they were concurrently engaged in distracting tasks during the game. Conversely, (Cabrales, Espín et al., 2022) reported that time pressure, whether immediate or delayed decision-making, did not significantly influence the decisions of trustors.

Moreover, (Krueger, Grafman et al., 2008), using simultaneous brain imaging of two participants engaged in a repeated trust game, demonstrated that distinct neural processes underlie the phases of trust formation and trust maintenance. Specifically, the paracingulate cortex in the prefrontal region was activated during initial trust formation, while the insula and septal areas were engaged during later stages of trust preservation.

Evolutionary models have also made substantial contributions to understanding the dynamics of trust. (Nowak, 2006) proposed five fundamental mechanisms for the evolution of cooperation: kin selection, direct reciprocity, indirect reciprocity (via reputation), network reciprocity, and group selection. Building on this framework, (Manapat, Nowak et al., 2013) demonstrated that partner selection can promote fair outcomes and support trust. Similarly, (Robert, 1984) showed that the Tit-for-Tat strategy facilitates stable cooperation in repeated games. In this context, (Cochard, Van et al., 2004) reported that in repeated versions of the trust game, both trust-giving and return rates increase over time, although a sudden drop in trust often occurs toward the end of finitely repeated games—a phenomenon known as the end-game effect. On a related note, (Fehr, 2009) argued that trust is not merely a form of risk-taking, but is instead grounded in social preferences, such as betrayal aversion. In a similar vein, (Bohnet and Zeckhauser, 2004) introduced the concept of betrayal aversion and showed that individuals require a significantly higher probability of return when trusting other humans than they do in equivalent random lotteries.

The role of reputation as a stabilizing mechanism in the dynamics of trust has been extensively documented within the framework of game theory. (Nowak and Sigmund, 1998) developed the theory of indirect reciprocity, suggesting that individuals are more inclined to trust those who possess a positive reputation history. This proposition has been empirically supported by (Weigelt and Camerer, 1988) as well as (Charness, Rigotti et al., 2016). Cross-cultural investigations, including those by (Croson and Buchan, 1999), further reveal gender and cultural variations in trust behavior and reciprocal responses.

The dynamics of trust in human–machine interactions have also become a significant area of research. (Zörner, Arts et al., 2021) demonstrated that nonverbal behaviors exhibited by robots, such as gesturing or verbal communication, enhance human trust in robotic agents. Similarly, (Aimone, Houser et al., 2014) found that individuals tend to trust less when they believe they are interacting with a computer compared to a human counterpart, aligning with the concept of betrayal aversion; however, the presence of a real human can mitigate this effect. In another study, (Han, Perret et al., 2021) showed that trust-based strategies perform better when interacting with artificial agents. (Dasgupta, 2009) and (Guiso, Sapienza et al., 2008) argued that trust improves economic performance and facilitates participation in financial markets. (Robles and Mallinson, 2023) emphasized that public trust is vital for the legitimacy of AI governance frameworks.

Despite the notable explanatory power of this model in capturing the dynamics of trusting behavior, it is important to recognize that the framework primarily focuses on the first player’s decision to initiate a trust-based interaction, while largely neglecting a systematic analysis of the strategies that the first player may employ to actively elicit trust from the counterpart. In much of the existing literature, the second player’s behavior is treated as a passive response to the initial move, and the act of trust attraction has received limited theoretical and empirical attention as an independent strategic action. This conceptual gap motivates the development of a novel approach in the present study, which, by centering on strategic actions aimed at eliciting trust, seeks to offer a complementary perspective to the classical trust game model.

Before turning to the core argument, it is essential to review the key factors that influence both the decision to trust and the ability to elicit trust—factors that have been examined in previous theoretical and empirical research. A broad body of literature has sought to identify the variables that lead individuals to accept the risk inherent in trusting—that is, to engage in trusting behavior with the expectation that the counterpart will respond in a trustworthy manner. Among the most prominent of these factors are the following:

Key factors identified in prior research include social preferences and fairness concerns (Fehr and Schmidt, 1999), feelings of guilt and moral commitments (Charness and Dufwenberg, 2006), biological influences such as the role of oxytocin (Kosfeld, Heinrichs et al., 2005), betrayal aversion and risk attitudes in human interactions (Bohnet and Zeckhauser, 2004), and cultural and gender differences (Croson and Buchan, 1999). A substantial body of experimental evidence confirms that trustors (i.e., first movers) do not base their decisions solely on material self-interest. Rather, normative, fairness-based, and social motivations play a significant role. (Fehr and Gächter, 2002) argued that individuals are willing to incur costs to prevent violations of social norms—a behavior known as altruistic punishment—which in turn fosters trust within social environments. The ERC (Equity, Reciprocity, and Competition) model proposed by (Bolton and Ockenfels, 2000) further highlights the relevance of sensitivity to relative payoffs in explaining cooperation and trust.
In line with this perspective, (Ashraf, Bohnet et al., 2006), in a cross-national experimental study conducted in three countries, found that the decision to trust is primarily driven by expectations of return on investment, whereas the degree of trustworthiness is more strongly influenced by unconditional benevolence—namely, the individual’s willingness to return amounts exceeding the expected level. Similarly, (Rodrigo-González, Caballer-Tarazona et al., 2021) showed that economic inequality can significantly affect trust: individuals tend to trust more those with lower past incomes or those who exert greater current effort. However, differences in relative wealth appear to exert a stronger influence on trustworthiness than on the initial willingness to trust.

In contrast, research on eliciting trust remains relatively limited. Nevertheless, several important findings from the fields of social psychology, organizational behavior, and communication management have identified key factors that contribute to this process. Warmth and behavioral simplicity, as opposed to displays of power, have been found to be more effective in generating trustworthiness perceptions (Cuddy, Kohut et al., 2013). Moreover, social signals conveyed through facial expressions and body language significantly shape initial trust impressions (Willis, Palermo et al., 2011). (Bacharach, Guerra et al., 2007) also demonstrated that early-stage signaling in the onset of an interaction can substantially increase the likelihood of reciprocal trust.

The present study shares strong thematic, conceptual, and methodological affinities with the work of (Bacharach, Guerra et al., 2007). Their experimental research demonstrated that sending honest signals and engaging in norm-consistent behavior during the early stages of interaction significantly increases the likelihood of reciprocal trust. Building on the trajectory established by their study, the current research seeks to develop a theoretical and mathematical framework for analyzing the process of trust elicitation within repeated interaction settings. This framework is grounded not only in the logic of repeated games but also draws inspiration from the reputation-based model proposed by (Sabater and Sierra, 2005), in which agents rely on structured information and prior evaluations to build and reinforce trust in multi-agent systems. Accordingly, the theoretical innovation of this paper lies in its integration of classical models of human signaling with algorithmic approaches designed for inter-agent interactions.

In addition to the issues discussed earlier, a review of further studies has revealed additional insights that merit inclusion at this stage. Presenting these findings here contributes to a more precise formulation of the research problem and underscores the theoretical and practical significance of the topic.

One of the well-documented strategies in intelligence operations and espionage is the use of deliberate behavioral signaling to elicit trust. For example, the British intelligence agency MI6 is known to have employed the tactic of feigning simplicity or harmlessness as part of the broader Gray Man Tactic. This approach involves intentionally minimizing one’s visibility in public spaces in order to reduce perceived threat levels. The individual acts in ways that appear entirely ordinary, inconspicuous, and sometimes even naïve or unintelligent. The core purpose of this strategy is to build trust, or at the very least, to neutralize initial suspicion in intelligence or field interactions.

(Schneier and Schneier, 2003), in his work on security behavior, explains how displaying non-threatening behavior and blending into the social environment can facilitate trust-building, particularly in urban security contexts and threat detection scenarios. He emphasizes that being overlooked can serve as an entry point for approaching a target without provoking defensive reactions.

Further studies on trust elicitation have examined the strategic use of deliberate societal fragmentation as a tool in cognitive and information warfare. One particularly effective method for gaining the trust of specific segments within a target society involves the creation or exacerbation of internal divisions—whether ethnic, cultural, generational, or socioeconomic. This tactic, commonly referred to as Divide and Trust, operates by undermining internal cohesion and redirecting trust toward an external actor.

(Paul and Matthews, 2016), in their strategic report on Russia’s Firehose of Falsehood propaganda model, describe how the simultaneous dissemination of contradictory information can foster confusion and internal distrust. Over time, this state of disorientation may lead segments of the population to place their trust in the external actor as a perceived stabilizing force. The model is especially effective in disrupting societal coherence and positioning the external communicator as a credible alternative.

Similarly, (Pomerantsev, 2019), in his influential work, illustrates how modern propaganda techniques—especially in the digital sphere—are designed to weaken internal trust and promote new narratives through the amplification of social cleavages. He analyzes the information strategies employed by state actors in countries such as Russia, China, and even liberal democracies.
In the field of political marketing, (Baines and O'Shaughnessy, 2014) argue that the strategic erosion of trust between social groups—such as between generations, ethnic communities, or economic classes—can effectively redirect allegiance toward a political message or alternative brand. Psychologically, this approach is conceptualized as targeted polarization, wherein trust is not gained through overt honesty but rather through the manipulation of perception, either by reducing perceived threat or by exploiting identity divisions.

Despite significant advances in understanding the dynamics of trust, a fundamental conceptual gap remains: how can an agent systematically and repeatedly elicit trust from a counterpart during the early stages of interaction through structured behavioral strategies? This question has been largely overlooked in the classical literature and, more specifically, lacks clear mathematical modeling within the framework of game-theoretic analysis.

Moreover, while the classical Trust Game primarily focuses on present-moment decision-making and the immediate outcomes of interaction, the Trust Attraction Game proposed in this study shifts attention to the future temporal dimension. In this framework, current actions are strategically designed to shape the counterpart’s expectations and future decisions, with outcomes that manifest over time rather than instantaneously.

Inspired by studies such as (Bacharach, Guerra et al., 2007), which demonstrate that early signaling of normative and honest behavior can foster the emergence of trust, the present study aims to formally, theoretically, and mathematically formulate this process as a strategic Trust Attraction Game.

Our proposed framework provides a rigorous modeling of the first player’s actions in the early stages of interaction, offering precise definitions of decision variables, counterpart beliefs, and the dynamics of trust updating. By extending the concept of initial signaling, we illustrate how the first player can foster mutual trust through behavioral consistency, gradual alignment, and the strategic selection of actions—even in the absence of prior familiarity or external guarantees.

From this perspective, the present paper addresses the existing theoretical gap by enabling formal analysis of the dynamics of trust elicitation within the framework of repeated games. It opens new avenues for studying strategies of social influence, political communication, and the design of participatory institutions. The Trust Attraction Game possesses broad applicability across diverse contexts, including familial interactions such as courtship processes, social and professional settings like recruitment and selection procedures, and even broader arenas such as international relations where one government seeks to gain the trust of another, or domestic politics involving the cultivation of public trust by a government toward its citizens.

\subsection{Example: Brand--Customer Relationship}

A concrete example of the Trust Game in practice is seen in the relationship between brands and their customers. When a new brand enters the market, consumers typically have little to no information regarding the quality of its products, which may lead to hesitation in purchasing. In this scenario, the brand must make initial efforts to demonstrate trustworthiness. These efforts may include offering guarantees, providing evidence of quality, or leveraging positive feedback from previous customers. These actions serve as signals that help reduce uncertainty and initiate a trust-building process.

As customers engage with the brand and experience positive outcomes, trust deepens, leading to a continued relationship. Over time, as the brand consistently meets customer expectations, the trust built through these repeated interactions can result in customer loyalty. Furthermore, satisfied customers may recommend the brand to others, thus expanding its reputation and fostering further trust.



\subsection{The Present Study: The Trust Attraction Game}

Building on these foundational concepts, the present study introduces the Trust Attraction Game, a novel framework designed to analyze the sustainability of trust in social and economic interactions. This model shifts the focus to incentivizing the second player's cooperation and gradually reinforcing trust through a series of repeated interactions. Unlike traditional models where trust is often seen as a static exchange, the Trust Attraction Game views trust as a dynamic, evolving process, where the second player’s decision-making is shaped by the consistency and reliability of the first player’s actions over time.

This framework offers a valuable lens through which we can examine how trust is cultivated and maintained in various settings, including business, politics, and international diplomacy. By providing a systematic approach to analyzing trust dynamics, the model can also inform the development of strategies for promoting sustainable cooperation across different sectors.

For further readings on trust dynamics and cooperation in game theory, the following studies offer valuable insights:

\begin{itemize}
    \item Liu, L., \& Chen, X. (2022). A conditional investment strategy in evolutionary trust games with repeated group interactions.
    \item Carlsson, F., Demeke, E., Martinsson, P., \& Tesemma, T. (2018). Measuring trust in institutions.
    \item Bayat, D., Mohamadpour, H., Fang, H., Xu, P., \& Krueger, F. (2023). The impact of order effects on the framing of trust and reciprocity behaviors.
    \item Rashid, M. M., Xiang, Y., Uddin, M. P., Tang, J., Sood, K., \& Gao, L. (2025). Trustworthy and fair federated learning via reputation-based consensus and adaptive incentives.
\end{itemize}

These studies contribute to a broader understanding of trust in game-theoretic models, focusing on reputation, institutional trust, and strategic interactions. Through the integration of these insights, this study aims to advance the theoretical discourse on trust and cooperation in complex social and economic systems.

These studies contribute to a deeper understanding of the factors shaping trust, including reputation, institutional trust, and strategic interactions within game-theoretic frameworks.


\subsection{The Trust Game}

The Trust Game is structured as follows:
\begin{itemize}
    \item Player 1 (Trustor) decides to send an amount $x$.
    \item The money received by Player 2 (Trustee) is multiplied by a factor $k$.
    \item Player 2 decides how much of the amount $kx$ to return.
\end{itemize}

Cooperation and trust are reinforced based on repeated interactions.

\begin{table}[htbp]
\centering
\caption{Payoff Table of the Trust Game}
\label{tab:trust_game}
\begin{tabularx}{\textwidth}{C C C C}
\toprule
\textbf{Player 1 (Trustor)} &
\textbf{Player 2 (Trustee)} &
\textbf{Payoff for Player 1} &
\textbf{Payoff for Player 2} \\
\midrule
Send $x$ & Returns $kx$ & $+\,$Profit & $+\,$Profit \\
Send $x$ & Does Not Return & $-x$ (Loss) & $+x$ \\
Does Not Send & -- & $0$ & $0$ \\
\bottomrule
\end{tabularx}
\end{table}

Below is the conceptual representation of the game tree: Player~1 begins by choosing whether to send money or not. Then, Player~2 decides whether to return the amount $kx$ or to keep it.

\begin{figure}[H]
\centering
\includegraphics[width=\textwidth]{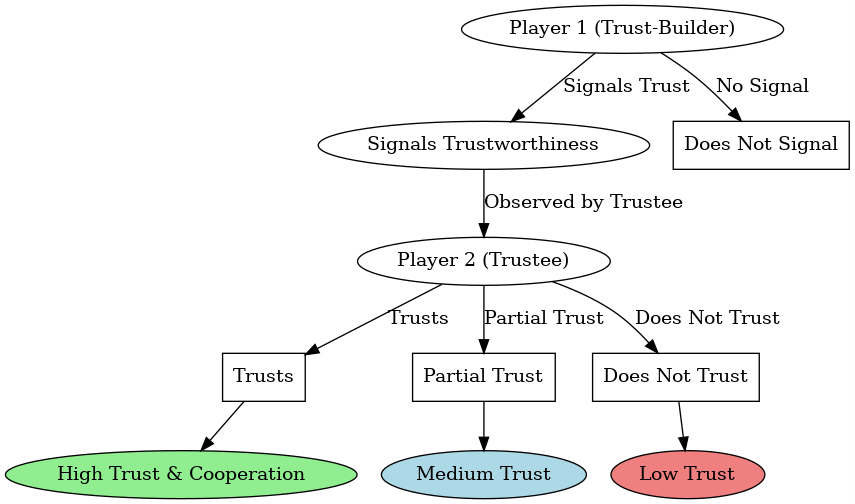}
\caption{Tree representation of the Trust Game.}
\label{fig:trust-tree}
\end{figure}


\section{Materials and Methods}

\subsection{The Trust-Building Game}

Unlike the classic Trust Game, in which Player 1 (the trustor) initiates trust by transferring part of their capital to Player 2 (the trustee), the Trust-Building Game introduced in this paper focuses on a more dynamic and realistic process of trust formation over repeated interactions. In this model, Player 1, referred to as the ``Trust Builder,'' must not only take the initial risk but also engage in deliberate and consistent signaling behavior to gradually earn the trust of Player 2. This reflects real-world scenarios more accurately, where trust typically emerges over time rather than in a single transaction.

The structure of the game unfolds as follows:

\begin{itemize}
    \item In each round, Player 1 (Trust Builder) performs an action designed to signal their trustworthiness. Such actions may include offering guarantees, providing unilateral benefits, sharing transparent information, or initiating cooperative behavior without immediate returns.
    \item Player 2 (Trustee) observes this action and decides whether to reciprocate by granting trust to Player 1.
    \item If Player 2 decides to trust, the interaction continues, allowing for the possibility of mutual long-term benefits. If Player 2 does not trust, the interaction either terminates or proceeds at a lower level of cooperation.
\end{itemize}

The trust level in this game is dynamic: it accumulates or deteriorates over time depending on Player 1’s behavior. A single misstep can significantly erode trust, even if it has been gradually building up over previous rounds.

The innovation of this model lies in its treatment of trust not as a one-shot transaction, but as a strategic, time-dependent process. In many real-life contexts—such as relationships between a customer and a brand, citizens and governments, or business partners—trust is rarely established instantly. Rather, it develops over time through repeated, observable behavior and perceived consistency.

We refer to this model as the \textit{Trust Attraction Game}, emphasizing the active role of Player 1 in attracting and maintaining trust. This formulation provides a novel conceptual tool for analyzing the dynamics of trust in both theoretical and applied game-theoretic settings, especially where long-term cooperation is essential.


\begin{table}[htbp]
\centering
\caption{Payoff Table of the Trust-Building Game}
\begin{tabularx}{\textwidth}{C C C C}
\toprule
\textbf{Player 1 (Trust-Builder)} &
\textbf{Player 2 (Trustee)} &
\textbf{Payoff for Player 1} &
\textbf{Payoff for Player 2} \\
\midrule
Signals Trustworthiness & Trusts & + Long-term Gain & + Cooperation Benefits \\
Signals Trustworthiness & Does Not Trust & - Initial Cost & 0 \\
Does Not Signal & -- & 0 & 0 \\
\bottomrule
\end{tabularx}
\end{table}

This game extends over multiple rounds, with trust gradually increasing or decreasing based on Player 1’s consistent or inconsistent behavior.

\begin{figure}[H]
\centering
\includegraphics[width=0.60\textwidth]{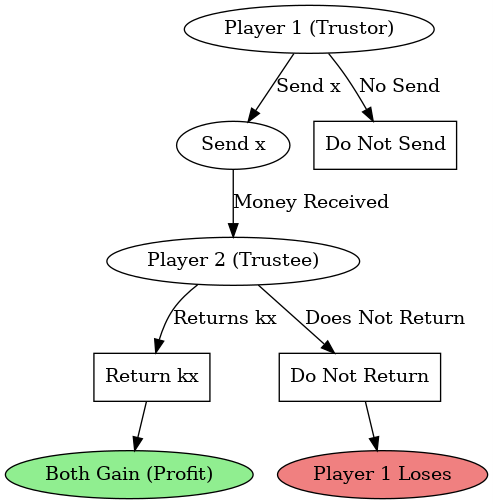}
\caption{ Game Tree for the Trust-Building Game}
\label{fig:trust-building-tree}
\end{figure}

\subsubsection{Equilibrium Analysis}

To identify the optimal strategic behavior in the Trust Attraction Game, we analyze its Nash Equilibrium in two key scenarios. This allows us to understand the conditions under which cooperation becomes a stable and self-enforcing outcome.

\paragraph{Case 1: The Trustee Trusts.}

In this favorable scenario, Player~2 (the Trustee) chooses to place trust in Player~1 (the Trust Builder) after observing consistent and credible signals of trustworthiness. As a result, cooperation between the two players is sustained across multiple rounds, enabling both parties to enjoy cumulative long-term benefits.

The Nash Equilibrium in this case is established when Player~1 finds it optimal to continue signaling trustworthiness because any deviation would risk the collapse of cooperation, and when Player~2’s best response is to reciprocate trust based on observed behavior. The mutual best responses reinforce each other, resulting in a stable cooperative relationship. This equilibrium reflects real-world trust scenarios where relationships flourish through mutual reinforcement and sustained positive interaction.

\paragraph{Case 2: The Trustee Does Not Trust.}

In contrast, if Player~2 withholds trust—perhaps due to insufficient or inconsistent signaling by Player~1—then cooperation fails to take root. Despite Player~1’s efforts, the absence of trust from Player~2 leads to a breakdown in interaction. In this setting, the Nash Equilibrium results in no cooperation, and Player~1 bears the cost of unreciprocated signaling.

These two cases highlight a critical insight: trust must be actively earned and strategically maintained. Player~1 must adopt a forward-looking strategy that builds a reputation for reliability. Without credible and consistent signals, trust cannot emerge, and the system defaults to a non-cooperative equilibrium—an outcome both inefficient and undesirable.


\subsubsection{Mathematical Modeling}

In this subsection, we develop a formal and fully mathematical framework for the Trust Attraction Game. The structure of the model is consistent with the classical reputation literature (Fudenberg and Levine, 1992; Kreps and Wilson, 1982; Mailath and Samuelson, 2006), but it differs from these models in both its objective and its state variable. Instead of modeling reputation about a hidden type, the present framework captures behavior in terms of a dynamic process of trust attraction. This distinction will be made precise in what follows.

\paragraph{Players, Time, and Information.}

The game is played by two agents:
\begin{itemize}
    \item \textbf{Player 1 (Trust Builder)}, who chooses publicly observable actions;
    \item \textbf{Player 2 (Trust Decision-Maker)}, who decides whether to grant trust based on the observed past behavior of Player~1.
\end{itemize}

Time is discrete:
\[
t = 0,1,2,\ldots
\]
and the horizon is infinite. Future payoffs are discounted using a common factor
\[
\delta \in (0,1).
\]

At the beginning of each period, both players observe the full history of Player~1's past actions. Let the publicly observed history up to period \(t\) be denoted by
\[
h_t = (s_0, s_1, \ldots, s_{t-1}),
\]
where \(s_\tau \in \{0,1\}\) is the action chosen by Player~1 in period \(\tau\).  
Players also observe the current value of the trust variable \(T_t\).  
Therefore, the information structure is one of \textbf{complete and public information}, with no private signals or hidden types.


\paragraph{State Variable: Trust as a Dynamic Behavioral Measure.}

Trust serves as the central state variable summarizing Player~2's assessment of Player~1's reliability based solely on observable behavior.  
The trust level at time \(t\) is represented by a continuous variable
\[
T_t \in [0,1],
\]
where \(T_t = 0\) denotes complete distrust and \(T_t = 1\) denotes full trust.

In contrast to classical reputation models—where the state variable typically reflects a belief about a fixed and unobservable type—Player~1 in the present framework has no hidden type, and all relevant aspects of behavior are fully observable.  
Accordingly, \(T_t\) does not represent a Bayesian belief about an underlying parameter; rather, it is a deterministic, observable summary of accumulated behavioral evidence.  

Trust evolves over time in response to Player~1's actions, increasing with consistent reliability and decreasing when actions provide evidence of inconsistency or reduced trustworthiness.

Accordingly, trust evolves dynamically over time: it increases when Player 1 behaves consistently and reliably, and it decreases when Player 1’s behavior signals inconsistency or unreliability. The trust variable therefore operates as a deterministic, observable state summarizing the behavioral history relevant for decision-making.

\paragraph{Dynamics of Trust.}

Given that trust operates as the publicly observed state variable of the game, its evolution is fully determined by Player~1's observable behavior. In each period \(t\), Player~1 chooses an action
\[
s_t \in \{0,1\},
\]
where \(s_t = 1\) denotes the emission of a positive, trust-building signal and \(s_t = 0\) denotes the absence of such a signal.

The trust level evolves according to the linear state-transition function
\[
T_{t+1} = \lambda T_t + (1-\lambda)s_t, \qquad \lambda \in [0,1],
\]
where the parameter \(\lambda\) captures the persistence of past trust. Higher values of \(\lambda\) assign greater weight to historical behavior, whereas lower values make current actions more influential in determining next-period trust.

\medskip
\noindent\textbf{Lemma (Boundedness of Trust).}  
If \(T_0 \in [0,1]\), then \(T_t \in [0,1]\) for all \(t \ge 0\).

\emph{Proof.}  
Since \(T_{t+1}\) is a convex combination of two values in the interval \([0,1]\), it must also lie in \([0,1]\). \(\square\)

\paragraph{Decision Rule of Player 2.}

Player~2 uses a threshold-based decision rule by comparing the current trust level $T_t$ with a fixed parameter $\theta \in [0,1]$, which represents the minimum level of trust required to justify a cooperative response. The action in period $t$ is given by
\[
a_t =
\begin{cases}
1, & \text{if } T_t \geq \theta,\\[4pt]
0, & \text{if } T_t < \theta.
\end{cases}
\]
Thus, Player~2 chooses to trust if and only if the observed level of trustworthiness meets or exceeds the threshold $\theta$. This captures a behavioral mechanism in which trust is granted only when sufficient positive evidence has accumulated in the state variable $T_t$.

\medskip
\noindent\textbf{Proposition (Optimality of the Threshold Rule).}
Suppose that:
\begin{itemize}
    \item Player~2 receives a strictly positive one-period payoff only when $(s_t,a_t) = (1,1)$, and zero otherwise;
    \item the trust variable evolves according to the linear update rule
    \[
    T_{t+1} = \lambda T_t + (1-\lambda)s_t, \qquad \lambda \in [0,1],
    \]
    as defined in the previous subsection;
    \item strategies are Markovian, so that $a_t$ depends on the current state $T_t$ only.
\end{itemize}
Then the threshold rule above constitutes a Markov best response of Player~2. In particular, there exists a cutoff $\theta^\ast \in [0,1]$ such that it is optimal for Player~2 to choose $a_t = 1$ if $T_t \geq \theta^\ast$ and $a_t = 0$ otherwise.

\emph{Sketch of proof.}
Given a Markov strategy $\sigma_1$ for Player~1, the induced process $\{T_t\}$ is Markovian with state space $[0,1]$, and the one-period payoff of Player~2 is nonnegative and strictly positive only when $(s_t,a_t)=(1,1)$. The continuation value of Player~2, denoted by $V_2(T)$, is nondecreasing in $T$ because higher trust levels weakly increase both the current probability of receiving the positive payoff $R_2$ and the expected future value (due to the monotone transition of $T_{t+1}$ in $T_t$). For each state $T$, Player~2 chooses between $a=0$ and $a=1$, where the difference in value,
\[
\Delta(T) \coloneqq \bigl[ u_2(\sigma_1(T),1) + \delta V_2(T') \bigr] 
- \bigl[ u_2(\sigma_1(T),0) + \delta V_2(\tilde T') \bigr],
\]
is monotone in $T$ (here $T'$ and $\tilde T'$ denote the next-period trust levels under $a=1$ and $a=0$, respectively). Monotonicity implies that the set of states for which $a=1$ is optimal is an interval of the form $[\theta^\ast,1]$, which yields a cutoff (threshold) policy. This structure is consistent with standard results on monotone optimal policies in dynamic control and Markov decision processes (see, e.g., Puterman, 1994). \hfill$\square$

\medskip
\paragraph{Payoff Functions.}

The one-period payoff of Player~1 is given by the function
\[
u_1(s_t,a_t) =
\begin{cases}
R_1 - c, & \text{if } s_t = 1,\ a_t = 1,\\[4pt]
-\,c, & \text{if } s_t = 1,\ a_t = 0,\\[4pt]
0, & \text{if } s_t = 0,\ a_t \in \{0,1\},
\end{cases}
\]
where $c>0$ is the cost of sending a trust-building signal and $R_1>0$ is the long-term benefit obtained when cooperation is successfully established.

The one-period payoff of Player~2 is
\[
u_2(s_t,a_t) =
\begin{cases}
R_2, & \text{if } s_t = 1,\ a_t = 1,\\[4pt]
0, & \text{otherwise},
\end{cases}
\]
where $R_2>0$ is the benefit that Player~2 receives when mutual cooperation occurs. We impose the natural parameter restrictions
\[
R_1 > c > 0, \qquad R_2 > 0,
\]
so that it is costly for Player~1 to send a signal, but sending a signal can be profitable in expectation when it leads to sustained cooperation.

\medskip
\paragraph{Markov Strategies.}

Because the trust variable $T_t$ serves as a sufficient statistic summarizing the entire history of past interactions, it is without loss of generality to restrict attention to Markov (or stationary) strategies that depend only on the current state. A Markov strategy for Player~$i$ is a mapping
\[
\sigma_i : [0,1] \to \{0,1\}, \qquad i = 1,2,
\]
where $\sigma_1(T)$ denotes the signal choice of Player~1 when the current trust level is $T$, and $\sigma_2(T)$ denotes the trust decision of Player~2 in the same state.

\medskip
\paragraph{Bellman Equations and Markov Perfect Equilibrium.}

Given a pair of Markov strategies $(\sigma_1,\sigma_2)$, the value function of Player~1 satisfies the Bellman equation
\[
V_1(T) = \max_{s \in \{0,1\}} 
\Bigl\{ u_1\bigl(s,\sigma_2(T)\bigr)
+ \delta\, V_1\bigl( \lambda T + (1-\lambda)s \bigr) \Bigr\},
\]
for all $T \in [0,1]$. Similarly, the value function of Player~2 is given by
\[
V_2(T) = \max_{a \in \{0,1\}} 
\Bigl\{ u_2\bigl(\sigma_1(T),a\bigr)
+ \delta\, V_2\bigl( \lambda T + (1-\lambda)\sigma_1(T) \bigr) \Bigr\},
\]
for all $T \in [0,1]$.

\medskip
\noindent\textbf{Definition (Markov Perfect Equilibrium).}
A pair of Markov strategies $(\sigma_1^\ast,\sigma_2^\ast)$ is a Markov Perfect Equilibrium (MPE) of the Trust Attraction Game if:
\begin{itemize}
    \item for each $T \in [0,1]$, $\sigma_1^\ast(T)$ attains the maximum in the Bellman equation for $V_1$ given $\sigma_2^\ast$;
    \item for each $T \in [0,1]$, $\sigma_2^\ast(T)$ attains the maximum in the Bellman equation for $V_2$ given $\sigma_1^\ast$;
    \item the associated value functions $(V_1,V_2)$ solve their respective Bellman equations and constitute a fixed point of the corresponding Bellman operators.
\end{itemize}

\medskip
\noindent\textbf{Proposition (Existence of Markov Perfect Equilibrium).}
In the Trust Attraction Game, there exists at least one Markov Perfect Equilibrium in (pure) Markov strategies.

\emph{Sketch of proof.}
The state space $[0,1]$ is compact, and the action sets for both players are finite, $\{0,1\}$. The one-period payoff functions $u_1$ and $u_2$ are bounded and continuous in $T$, and the state transition
\[
T_{t+1} = \lambda T_t + (1-\lambda)s_t
\]
is continuous in $(T_t,s_t)$. For any fixed pair of Markov strategies $(\sigma_1,\sigma_2)$, the induced Bellman operators for $V_1$ and $V_2$ are contractions on the space of bounded continuous functions on $[0,1]$ under the supremum norm, due to the discount factor $\delta \in (0,1)$. Hence, for each strategy profile $(\sigma_1,\sigma_2)$, the associated value functions $(V_1,V_2)$ exist and are unique.

The set of Markov strategies can be identified with the set of measurable mappings from the compact state space $[0,1]$ into the finite action sets, which is itself compact under the product topology. The best-response correspondence that maps a Markov strategy of one player into the nonempty, compact set of best-response Markov strategies of the other player is upper hemicontinuous. By Kakutani's fixed point theorem, there exists a fixed point of the joint best-response correspondence, i.e.\ a pair $(\sigma_1^\ast,\sigma_2^\ast)$ such that each is a best response to the other. Such a pair is a Markov Perfect Equilibrium. This argument is standard in the theory of dynamic games with compact state space and finite action sets; see, for instance, Shapley (1953) and Stokey, Lucas, and Prescott (1989) for related existence results in stochastic and dynamic programming frameworks. \hfill$\square$

\paragraph{Optimal Behavior of Player 1.}

Because Player~2 follows a threshold rule with cutoff $\theta$ (as established in the previous subsection), Player~1's incentives depend entirely on whether the current trust level $T$ is sufficiently high for Player~2 to respond cooperatively. When $T < \theta$, sending a signal increases the future probability of cooperation; when $T \ge \theta$, cooperation is already ensured and further signaling only imposes unnecessary cost. This structure induces a natural monotonicity in Player~1's dynamic optimization problem.

\medskip
\noindent\textbf{Theorem 1 (Existence of an Optimal Trust-Building Threshold).}
Assume $R_1 > c > 0$. Then Player~1's optimal Markov strategy admits a cutoff $\widehat{T} \in [0,1]$ such that
\[
\sigma_1^\ast(T) =
\begin{cases}
1, & \text{if } T < \widehat{T},\\[4pt]
0, & \text{if } T \ge \widehat{T}.
\end{cases}
\]
That is, Player~1 finds it optimal to send costly trust-building signals only when the trust level is below a critical threshold; once trust becomes sufficiently high, further signaling is suboptimal.

\medskip
\noindent\emph{Sketch of proof.}
Under Markov strategies, Player~1 solves the Bellman equation
\[
V_1(T)=\max_{s \in \{0,1\}} \left\{
u_1\bigl(s,\sigma_2(T)\bigr)
+ \delta V_1\bigl( \lambda T + (1-\lambda)s \bigr)
\right\},
\]
and $\sigma_2(T)$ is a monotone threshold rule with cutoff $\theta$.  
The continuation value $V_1(T)$ is increasing in $T$ because higher trust levels weakly increase the probability of receiving the cooperative payoff $R_1$ in the future.  
The difference between the two Bellman branches,
\[
\Delta(T)
= \bigl[R_1-c + \delta V_1(T')\bigr]
- \bigl[0 + \delta V_1(\widetilde{T}')\bigr],
\]
where $T'$ and $\widetilde{T}'$ denote next-period trust levels under $s=1$ and $s=0$, is strictly decreasing in $T$ due to the transition rule $T_{t+1}=\lambda T_t + (1-\lambda)s_t$ and the monotonicity of $V_1$.  
Thus, the set of trust states in which signaling is optimal is an interval of the form $[0,\widehat{T})$.  
This single-crossing structure is a standard consequence of monotone dynamic programming (see Puterman, 1994). \hfill$\square$

\medskip
\paragraph{Formal Distinction from the Reputation Literature.}

\noindent\textbf{Theorem 2 (Non-Reducibility to Classical Reputation Models).}
The Trust Attraction Game cannot be transformed into a classical reputation model with a fixed hidden type. In particular, there exists no one-to-one mapping between the trust variable $T_t$ and any belief state $\mu_t$ that arises from Bayesian updating over a static type space.

\medskip
\noindent\textit{Proof (Conceptual).}
The distinction follows from five structural properties:
\begin{enumerate}
    \item \textbf{Dynamic behavioral state.}  
    The state variable $T_t$ represents a deterministic, continuous, and controllable behavioral variable, not a posterior belief over a fixed and unobservable type.
    
    \item \textbf{Controllability of behavior.}  
    Player~1 directly influences the evolution of $T_t$ via the choice of $s_t$.  
    In classical reputation models, by contrast, the type is exogenous and fixed.
    
    \item \textbf{Non-Bayesian updating.}  
    The law of motion $T_{t+1}=\lambda T_t+(1-\lambda)s_t$ is not derived from Bayes' rule. No likelihood function or prior enters the state transition.
    
    \item \textbf{Full information.}  
    Player~2 observes all actions and the state $T_t$ with no uncertainty, whereas reputation models require informational asymmetry.
    
    \item \textbf{No hidden type.}  
    Because the game contains no hidden parameter of Player~1, the Kreps–Wilson framework (1982), in which beliefs about hidden types drive reputation effects, does not apply.
\end{enumerate}
Therefore, the present model is structurally distinct from classical reputation models and cannot be reduced to a Bayesian belief-updating framework. \hfill$\square$

\section{Results}

By varying the memory parameter $\lambda$, we observe the following behavioral patterns: Psychological or social characteristics of individuals can be modeled through this parameter. For instance, customers who are quickly influenced by advertisements or brand-related behavior tend to exhibit a low $\lambda$. In contrast, individuals or nations that do not easily forget negative past experiences, even when faced with improved behavior from the other party, are characterized by a high $\lambda$.

Accordingly, trust-building strategies should be tailored based on the counterpart’s memory sensitivity:
\begin{itemize}
    \item \textbf{When the target audience has a low $\lambda$:} a single positive action may be sufficient to rapidly establish trust.
    \item \textbf{When the target audience has a high $\lambda$:} multiple consistent and reliable behaviors are required for trust to emerge.
\end{itemize}

The parameter $\lambda$ captures the degree to which Player~2’s decision making is shaped by recent versus cumulative experience. Specifically, low values of $\lambda$ reflect short-term memory in which recent interactions are heavily weighted, whereas high values reflect long-term memory, where earlier experiences continue to influence judgment.

It is also essential to recognize that trust formation translates into actual behavioral responses only when the trust level $T_t$ exceeds a predefined threshold $\theta$. In our model, this dynamic is represented by the decision rule:
\[
a_t =
\begin{cases}
1, & T_t \ge \theta,\\[4pt]
0, & T_t < \theta.
\end{cases}
\]

Thus, when $\lambda$ is low, trust rises more quickly, and Player~2 is likely to reach the threshold $\theta$ sooner, resulting in earlier trust-related actions (i.e., $a_t = 1$). Conversely, for high values of $\lambda$, trust accumulates more gradually, and it takes longer for $T_t$ to surpass the threshold, delaying the manifestation of trust behavior. This highlights the joint influence of both memory sensitivity and the trust threshold in shaping Player~2’s observable decisions.

The simulation in Figure~4 illustrates how the memory parameter $\lambda$ affects the speed of trust formation. A low value of $\lambda$ reflects short-term memory, where recent behaviors are weighted more heavily, leading to faster trust development. In contrast, a high $\lambda$ corresponds to long-term memory, where past behaviors dominate the decision-making process, resulting in delayed trust responses.

This insight underscores the importance of adapting trust-building strategies to the psychological characteristics of the counterpart, particularly in systems such as brand loyalty or diplomatic relations.
\begin{figure}[H]
\centering
\includegraphics[width=0.8\textwidth]{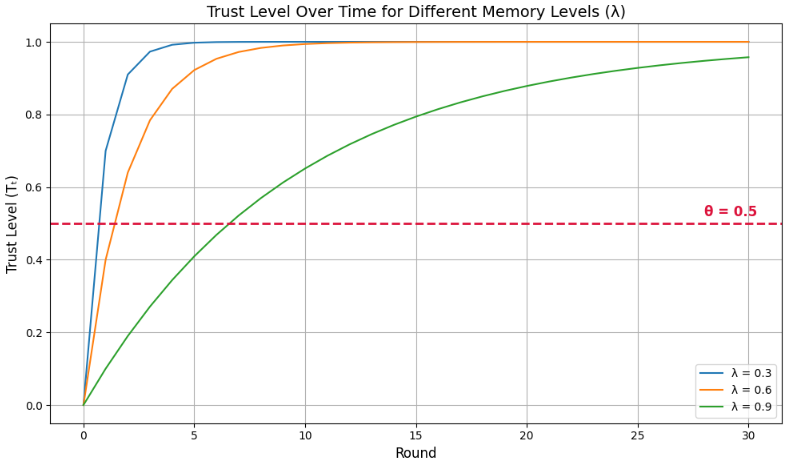}
\caption{Figure 3. Evolution of Player~2’s trust over time as a function of the trust memory parameter $\lambda$.}
\label{fig:trust_evolution}
\end{figure}
\begin{table}[htbp]
\centering
\caption{Table 4. Summary of trust levels across different $\lambda$ values}
\begin{tabularx}{\textwidth}{C C C C}
\toprule
\textbf{Memory Level ($\lambda$)} &
\textbf{Trust Formation Speed} &
\textbf{When Trust Crosses Threshold ($T_t \ge \theta$)} &
\textbf{Player~2's Behavior ($a_t$) After Threshold} \\
\midrule
Low ($\lambda = 0.3$) &
Fast, influenced by recent signals only &
Around Round 3 &
Trust is placed frequently and immediately ($a_t = 1$ in most rounds) \\
Medium ($\lambda = 0.6$) &
Moderate, balances past and recent behavior &
Around Round 5 &
Trust is placed consistently starting from round 5 ($a_t = 1$) \\
High ($\lambda = 0.9$) &
Slow, strongly shaped by long-term history &
Around Round 10 &
Trust is placed gradually in later rounds ($a_t = 1$ appears increasingly) \\
\bottomrule
\end{tabularx}
\end{table}

\subsection{Conclusion from the Simulation}

Three distinct values of the memory parameter $\lambda$ (0.3, 0.6, and 0.9) were employed to represent short-term, medium-term, and long-term memory, respectively. In each round, Player~2’s decision was determined by comparing the calculated trust level with a predefined trust threshold $\theta = 0.5$. If the trust level was equal to or greater than the threshold, Player~2 chose to trust ($a_t = 1$), resulting in the expression of trustful behavior. Otherwise ($a_t = 0$), Player~2 refrained from trusting, and no trust-based interaction was initiated.

In the presented figure, in addition to illustrating the trust formation process for each value of $\lambda$, Player~2’s decisions at each round are also displayed. Green dots indicate rounds in which Player~2 placed trust (i.e., $a_t = 1$), while red dots represent rounds in which Player~2 decided not to trust (i.e., $a_t = 0$).

The simultaneous display of these points on the graph enables an analysis of the dynamic relationship between the evolution of trust and Player~2’s behavioral responses throughout the game. In particular, it reveals how behavioral memory (captured by the value of $\lambda$) and the trust threshold $\theta$ jointly influence Player~2’s decision-making process.

According to the proposed model and simulation results, the memory parameter $\lambda$ plays a crucial role in the speed of trust formation. In the case of short-term memory (low values of $\lambda$), the trust level $T_t$ increases more rapidly and crosses the trust threshold $\theta$ earlier, resulting in trust decisions $a_t = 1$ occurring in the initial rounds and a higher proportion of such decisions within a limited time frame.

Conversely, with long-term memory (high values of $\lambda$), the trust level $T_t$ grows gradually and crosses the threshold $\theta$ at a later stage, leading to fewer rounds where trust decisions $a_t = 1$ are made within the same limited time period.

Therefore, higher values of $\lambda$ are associated with a relative decrease in the proportion of trust decisions $a_t = 1$ over shorter periods, a trend that is clearly observable in the heatmap. These findings align with the psychological concept of memory and sensitivity to past experiences, emphasizing the importance of tailoring trust-building strategies according to the memory characteristics of the counterpart.
\begin{figure}[H]
\centering
\includegraphics[width=0.8\textwidth]{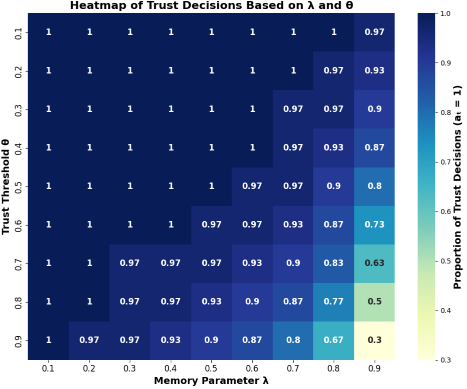}
\caption{Figure 4. Heatmap showing how trust decisions ($a_t = 1$) vary with the memory parameter $\lambda$ and the trust threshold $\theta$ over 30 rounds.}
\label{fig:heatmap_lambda_theta}
\end{figure}
\section{Conclusions}

This paper introduces the Trust-Building Game, which significantly expands upon the classic Trust Game by modeling trust from a different perspective. In the classic game, trust is initiated by Player~1 (the trustor) through a one-time transfer of resources to Player~2 (the trustee). However, in our proposed model, the focus shifts to Player~2. In this model, Player~1 (the trust-builder) must gradually earn Player~2's trust through repeated and strategic actions over time.

In this game, trust is a gradual process, built through consistent, reliable signals and actions by Player~1 over time. Player~2, responsible for deciding whether to trust Player~1, must base their decision on these signals and behaviors. This approach contrasts with the classic game, where Player~2 makes an immediate, one-shot decision regarding trust.

The findings suggest that trust in long-term interactions is strengthened through consistent and reliable behavior from Player~1, and this model can be applied in various fields such as brand–consumer relationships and public policy. The Trust-Building Game offers a more nuanced and realistic approach to understanding trust dynamics, highlighting the active role of Player~2 in determining the course of cooperation.

Future research could investigate practical applications in real-world contexts, such as:
\begin{itemize}
    \item Corporate collaborations and partnerships,
    \item International diplomacy and cross-border relations,
    \item Consumer trust in online transactions and e-commerce platforms.
\end{itemize}

By incorporating appropriate incentive structures, both organizations and policymakers have the potential to strengthen cooperation and develop sustainable, long-term strategies for trust-building across various domains.
\section*{References}
\begin{enumerate}

\item Berg, J., J. Dickhaut, and K. McCabe, Trust, reciprocity, and social history. \textit{Games and Economic Behavior}, 1995. 10(1): 122–142.

\item Johnson, N.D. and A. Mislin, How much should we trust the World Values Survey trust question? \textit{Economics Letters}, 2012. 116(2): 210–212.

\item Johnson, N.D. and A.A. Mislin, Trust games: A meta-analysis. \textit{Journal of Economic Psychology}, 2011. 32(5): 865–889.

\item Charness, G. and M. Dufwenberg, Promises and partnership. \textit{Econometrica}, 2006. 74(6): 1579–1601.

\item Zhang, M., D. Ji, and X. Chen, Building trust in participatory design to promote relational network for social innovation. In \textit{Proceedings of the Participatory Design Conference 2022 – Volume 2}, 2022.

\item Samson, K. and P. Kostyszyn, Effects of cognitive load on trusting behavior – an experiment using the trust game. \textit{PLOS ONE}, 2015. 10(5): e0127680.

\item Cabrales, A., et al., Trustors’ disregard for trustees deciding quickly or slowly in three experiments with time constraints. \textit{Scientific Reports}, 2022. 12(1): 12120.

\item Krueger, F., J. Grafman, and K. McCabe, Neural correlates of economic game playing. \textit{Philosophical Transactions of the Royal Society B}, 2008. 363(1511): 3859–3874.

\item Nowak, M.A., Five rules for the evolution of cooperation. \textit{Science}, 2006. 314(5805): 1560–1563.

\item Manapat, M.L., M.A. Nowak, and D.G. Rand, Information, irrationality, and the evolution of trust. \textit{Journal of Economic Behavior \& Organization}, 2013. 90: S57–S75.

\item Robert, A., \textit{The evolution of cooperation}. Basic Books, 1984.

\item Cochard, F., P.N. Van, and M. Willinger, Trusting behavior in a repeated investment game. \textit{Journal of Economic Behavior \& Organization}, 2004. 55(1): 31–44.

\item Fehr, E., On the economics and biology of trust. \textit{Journal of the European Economic Association}, 2009. 7(2–3): 235–266.

\item Bohnet, I. and R. Zeckhauser, Trust, risk and betrayal. \textit{Journal of Economic Behavior \& Organization}, 2004. 55(4): 467–484.

\item Nowak, M.A. and K. Sigmund, Evolution of indirect reciprocity by image scoring. \textit{Nature}, 1998. 393(6685): 573–577.

\item Weigelt, K. and C. Camerer, Reputation and corporate strategy. \textit{Strategic Management Journal}, 1988. 9(5): 443–454.

\item Charness, G., L. Rigotti, and A. Rustichini, Social surplus determines cooperation rates in the one-shot Prisoner's Dilemma. \textit{Games and Economic Behavior}, 2016. 100: 113–124.

\item Croson, R. and N. Buchan, Gender and culture: International experimental evidence from trust games. \textit{American Economic Review}, 1999. 89(2): 386–391.

\item Zörner, S., et al., An immersive investment game to study human-robot trust. \textit{Frontiers in Robotics and AI}, 2021. 8: 644529.

\item Aimone, J.A., D. Houser, and B. Weber, Neural signatures of betrayal aversion. \textit{Proceedings of the Royal Society B}, 2014. 281(1782): 20132127.

\item Han, T.A., C. Perret, and S.T. Powers, When to (or not to) trust intelligent machines. \textit{Cognitive Systems Research}, 2021. 68: 111–124.

\item Dasgupta, P., Trust and cooperation among economic agents. \textit{Philosophical Transactions of the Royal Society B}, 2009. 364(1533): 3301–3309.

\item Guiso, L., P. Sapienza, and L. Zingales, Trusting the stock market. \textit{The Journal of Finance}, 2008. 63(6): 2557–2600.

\item Robles, P. and D.J. Mallinson, Catching up with AI. \textit{Politics \& Policy}, 2023. 51(3): 355–372.

\item Fehr, E. and K.M. Schmidt, A theory of fairness, competition, and cooperation. \textit{Quarterly Journal of Economics}, 1999. 114(3): 817–868.

\item Kosfeld, M., et al., Oxytocin increases trust in humans. \textit{Nature}, 2005. 435(7042): 673–676.

\item Fehr, E. and S. Gächter, Altruistic punishment in humans. \textit{Nature}, 2002. 415(6868): 137–140.

\item Bolton, G.E. and A. Ockenfels, ERC: A theory of equity, reciprocity, and competition. \textit{American Economic Review}, 2000. 91(1): 166–193.

\item Ashraf, N., I. Bohnet, and N. Piankov, Decomposing trust and trustworthiness. \textit{Experimental Economics}, 2006. 9: 193–208.

\item Rodrigo-González, A., M. Caballer-Tarazona, and A. García-Gallego, Effects of inequality on trust and reciprocity. \textit{Frontiers in Psychology}, 2021. 12: 745948.

\item Cuddy, A.J.C., M. Kohut, and J. Neffinger, Connect, then lead. University of Washington, 2013.

\item Willis, M.L., R. Palermo, and D. Burke, Judging approachability on the face of it. \textit{Emotion}, 2011. 11(3): 514.

\item Bacharach, M., G. Guerra, and D.J. Zizzo, The self-fulfilling property of trust. \textit{Theory and Decision}, 2007. 63: 349–388.

\item Sabater, J. and C. Sierra, Review on computational trust and reputation models. \textit{Artificial Intelligence Review}, 2005. 24: 33–60.

\item Schneier, B. and B. Schneier, \textit{Beyond Fear}. Springer, 2003.

\item Paul, C. and M. Matthews, The Russian ``firehose of falsehood'' propaganda model. \textit{RAND Corporation}, 2016.

\item Pomerantsev, P., \textit{To Unreality—and Beyond}. 2019.

\item Baines, P.R. and N.J. O'Shaughnessy, \textit{Political Marketing and Propaganda}. Taylor \& Francis, 2014.

\end{enumerate}

\end{document}